# Self-assembly of Upconversion Nanoparticles and its luminescence

Monami Das Modak[c], Anil Kumar Chaudhary[b*] and Pradip Paik[a*]

[c]School of Engineering Sciences and Technology, University of Hyderabad (UoH), Hyderabad, TS, 500 046

[b]Advanced Centre of Research in High Energy Materials, University of Hyderabad (UoH), Hyderabad, TS, 500 046

[a]School of Biomedical Engineering, Indian Institute of Technology (IIT)-BHU, Varanasi, UP, 221 005

*Corresponding author's email: paik.bme@iitbhu.ac.in, pradip.paik@gmail.com, akcsp@uohyd.ernet.in, akcphys@gmail.com

**Abstract:**

We report on the in situ formation of 2D and 3D self assembled superlattices of upconverting nanoparticles. These as synthesized self assembled nanophosphors can emit sharp and intence luminescence and fluorescence using tunable wavelength femtosecond laser interaction and NIR 980 nm CW laser source, respectively. The relative up conversion luminescence intensities and the number of absorbed photons per photon emitted under the fs laser excitation power corresponds to each of the luminescence have been evaluated. The internal and external quantum yield of the self-assembled nanoparticles have also been expounded with different laser irradiations. All these results directed that a huge possible potential applications of these upconverting self-assembly materials that can make them significant in the electronic industry, such as for device making and for biomedical applications.

**Introduction:**

Since decades, self-assembled nanoparticles is having great importance in various fields including electronic device fabrication to the biomedical applications [1]. Self-assembled materials are available in nature; however, it can also be designed artificially by changing the different interactions between the nanoparticles and through varying the kinetic or the thermodynamic parameters in-situ. Biomaterials possess intrinsic behaviour and comprises self-assembled structure similar to the naturally occurring one [2]. However, self-assembly process is induced by the several factors and the mechanism of the formation fundamentally needs to be understood better to tune its properties [3–5]. Nanoparticle superlattice is one of the well-defined self-assembly ordered structure of nanoparticles and is different from the material of its bulk phase crystal, isolated nanocrystal and even disordered nanocrystal assemblies, which exhibits enhanced properties compared to its bulk like, conductivity, mechanical properties, optical properties, plasmonic properties, etc. depending on its order of packing and arrangement [6–10]. Nanoparticle superlattice can be constituted through the bottom-up self-assembly approach, where the process either can be in an equilibrium or in a non-equilibrium state. Usually, to achieve the highly order of nanoparticles in superlattice, the process involves soft ligands or capping agents, which further helps to tune the artificial structure and its properties through functionalizing the nanoparticles and by changing the particle-particles binding interactions. These soft ligands further can control the nearest-neighbour spacing, lattice structure and superlattice properties [6]. Thus, the collective interactions between organized superlattice nanoparticles are both of two dimension (2D) as well as three-dimension (3D), which make them effective for fabricating the electronic and optical devices [11–20]. It can be noted that the formation of superlattices of nanoparticles in large scale area is critical.

In 1994, Janos Fendler first introduced the concept of superlattice of nanoparticles [21,22] and produced CdS-nanoparticle (dia~ 26.5 Å to 34.0 Å) based superlattice [21] which was prepared at the air-water interface in a Langmuir film balance and the change in fluorescent properties was observed. In the same year, following the similar method, at air-water interface in a Langmuir film balance they constituted multi-layered superlattice of Ag nanoparticles (dia.~100 Å) with enhanced optical and electrooptical properties [22]. In subsequent year, 3D quantum dots (QDs) (CdSe, 20Å) superlattice structure (5 µm to 50 µm) was reported by Murray et al., where the spacing between the dots were controlled near the atomic scale precision [23]. They observed the discrete and size dependent optical absorption and band edge emission due to the quantized electronic transition of the individual QDs. Comparison to the optical properties between colloidal QDs and closed packed superlattice structure, the band spectra were identical. However, the shape of the emission spectra of the QDs in the superlattice was reformed and the red-shifting due to the inter-dots coupling was observed [23]. Superlattices of cobalt (Co) nanocrystal prepared in solution phase dispersion of Co nanoparticle using octane as solvent and oleic acid as capping agent with a nanocrystal spacing of ~ 4 nm which exhibited spin-dependent electron tunnelling properties [24]. In the similar line, nanoparticle superlattice alter the materials properties from its bulk, such as insulator-to-metallic transition [25] and leads to the enhancement of p-type conductivity (e.g., $PbTe/Ag_2Te$) [8], highly ordered vibrational coherency (e.g, FCC Ag NPs), plasmonic properties [7] etc. with retaining their basic solid-state crystal structure [6].

In this work, we report on the in-situ synthesis of upconversion nanoparticles and designed a number of highly ordered superlattices (UCN-SL) structures. Self-assembled UCN-SL is prepared by the one pot chemical synthesis approach. Their surface properties and upconversion and fluorescent properties have been studied using a continuous wave laser (CW-Laser) source and with a femtosecond laser (Fs) irradiation source.

It can be noted that, a huge number of potential applications of UCNPs make them significant in research due to their unique fluorescence and luminescence properties [26–30]. To the best of our knowledge, for the first time this work reports on the controlled synthesis for achieving shape and size directed UCN-SL having efficient fluorescence and luminescence properties under continuous wavelength (CW) source as well as femtosecond-laser (Fs) source.

Further, upconversion emissions at ultraviolet and visible emissions have been investigated. The synthesis procedure applied here to form UCN-SL is a modified approach of the synthesis method developed by our group previously. The entire reaction process was carried out in an inert gas (Argon) atmosphere. The different self-assembly superlattice structure have been formed in in-situ with varying the conditions. Usually researchers design this type of assembled structure in at least $2^{nd}$ stage of the whole synthesis procedure [1, 3–5], however, we are able to prepare such assembled structure in one-pot chemical synthesis approach. The beauty of our process is that the entire synthesis process consumes less time to obtain the self-assembled UCN-SL structures. Further, how the UCN-SL with different architecture exhibited different upconversion luminescence properties that also have been elucidated.  As-synthesized UCN-SL structures represented here are not only generate new physical properties compared to its bulk materials but also exhibits new phenomenon for interactions between particles at nanoscale. In order to produce a defined self-assembled UCN-SL pattern with synthesized nano-material, we have controlled the few reaction parameters. By applying those key factors, the correlative arrangements of nanocrystal building blocks with their sizes and the spacing between those crystals are maintained in a self-assembled long-range order. The precisely engineered nanocrystal's sizes and systematic shaped-assembly are treated as building blocks to generate the unique structures. For the UCN-SLs, the difference in the intensity of emission and their position and shifting, internal upconversion quantum yield (Internal UCQY) and external upconversion quantum yield (External UCQY), photoluminescence behaviour, fs-laser power dependent luminescence and the number of absorbed photons took part in the process, where the number of absorbed NIR photons (n) per emission under fs-laser excitation power has been calculated. The controlled morphology with efficient upconversion fluorescence and luminescence of these UCN-SLs may discover considerable applications in different fields such as for making adverse laser technology [31, 32], fabrication of diodes [33, 34], display and energy– devices [30, 35–42].

**2. Experimental Section of UC-SL:**

**2.1. *Materials*:**

$YCl_3:6H_2O$; $YbCl_3:6H_2O$ and $ErCl_3:6H_2O$ precursors (Sigma–Aldrich and with 99% purity), 1-Octadecene (90%), Oleic Acid (65%) NaOH (97.0%), $NH_4F$ (95 %) were purchased from sigma Aldrich, Qualigens, SDFCL and Kemphasol, respectively and were used for the synthesis of UCN-superlattices without any further purifications.

**2.2. *Synthesis of UCN-SL*:**

UCN-SLs were synthesized by thermal and solvothermal decomposition procedures of lanthanide hexahydrate precursors and technical grade chemicals The detail of synthesis procedure has been filled for Indian patent (Stable upconversion nanoparticle super- lattice (UCN-SL) & in-situ process for developing thereof; Ref. No./Application No.- 201841037607, App. Number: TEMP/E1/40924/2018- CHE; C.B.R. No.- 28247). In brief of the synthesis procedure: the specific amount of precursor materials ($YCl_3:6H_2O$; $YbCl_3:6H_2O$ and $ErCl_3:6H_2O$) were decomposed at ~ $120°$ C to remove moisture. In the next step, the organic solvents (15 ml 1-Octadecene and 6 ml Oleic-acid) were poured and stirred along with the decomposed components followed by heating at ~ $130°$ C -$140°$ C in Ar-gas environment. Thereafter, for removing the excess oxygen and moisture from the dissolved solution, a mixture (4:1) of NaOH and $NH_4F$ (dissolved in MeOH) was added at room temperature and continuously stirred for 1 h. The resulted solution was again heated under inert gas environment up to 300°C-350°C with a heating rate of 20 °C/min. At ~ $100°$ C the reaction chamber was degassed followed by purging of Ar gas and then heated up to $350°$ C. The whole system then cooled down to room temperature after 2 h of heat treatment procedure. Hence, in 1st set of experiments (Exset-1), the high reaction temperature was incorporated for 15-20 mins; in 2nd set of experiments (Exset-2), high temperature heat treatment was involved for about 50 mins. In 3rd set (Exset-3), high reaction temperature was maintained for about 1 h 50 mins. Next day, the synthesized nanoparticles were collected with acetone via high-speed centrifugation with 9000 RPM for few mins. The precipitated parts were gathered by dissolving with ~40 ml cyclohexane. Ethanol and $H_2O$ (1:1) was used for washing for 5-6 times. The resulted solutions were preserved as their colloidal form. After preserving, for the first couple of weeks, the colloidal solutions appeared without any agglomerations, but after a couple of months, the particles found to be settled at the bottom and can easily be dispersed in solution again.

**2. 3. Characterization technique**:

Transmission Electron Microscopy (TEM) and High-Resolution Transmission Electron Microscopy (HRTEM) (model FEI TecnaiG2-TWIN 200 KV) were used to analyse the size and morphology of the nanoparticles and of the as-synthesized UCN-SLs. Elemental analysis was carried out with Energy Dispersive X-ray Analysis (EDXA), determination of crystal structure was further carried out using X-Ray diffraction pattern (XRD with CoKα-radiation). Raman Spectroscopy (Wi-Tec, alpha 300) was performed to reveal the solid-state structure and defects present in UCN-SLs; Fluorescence Spectrophotometer (HITACHI, F-4600) was attached with a near infrared (NIR) laser source (980nm) externally to study the upconversion emissions. Further, a Ti-Sapphire tuneable oscillator was also used to study the upconversion luminescence behaviour along with femtosecond (fs) laser-pulses.

**Results and Discussion:**

As-synthesized ordered self-assembled UCN-SLs nanocrystal-building blocks are obtained according to the method as discussed in the experimental section. The colloidal chemistry and interactions played important role to self-organize the UCNPs into super lattice array within the colloidal solution. Formation of nano clusters also identified during the self-assembly arrangement. These clusters are made of with few monodispersed particles in a fascinating manner and it resulted into a periodic array of the particles to form the superlattices. The stability of the SL colloidal solutions can only be possible due to the presence of electrostatic interactions between the particles itself. As the nanoparticles are covered with ligands (oleic acid), they are dispersed transparently in solvent (cyclohexane). From the zeta potential results, it is confirmed that the UCN NPs are quite stable their values varied from -20 eV to -39.9 eV (Figures are not shown).

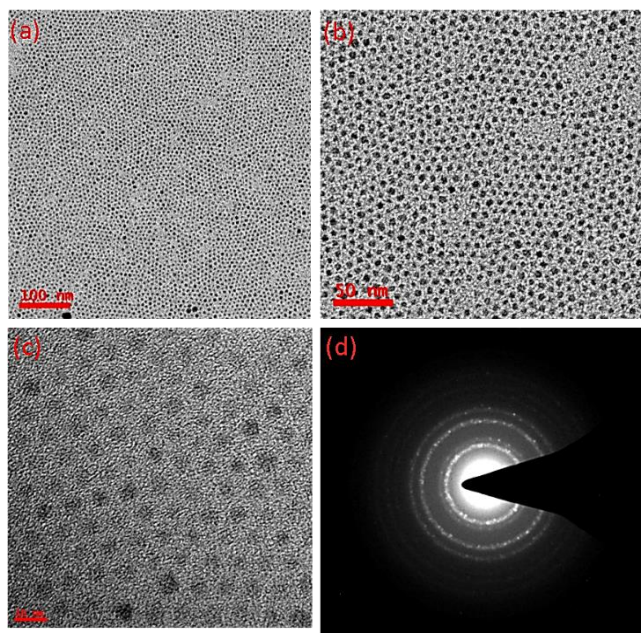

Fig.1. Superlattice pattern type 1: A uniform arrangement of synthesized UCNPs has been shown here forming the type-1 superlattice pattern. The distribution and formations of such a pattern is shown with different magnification (scale bar: 100 nm, 50 nm, and 10nm) in Fig (a), (b) and (c), respectively. SAED pattern is shown in (d).

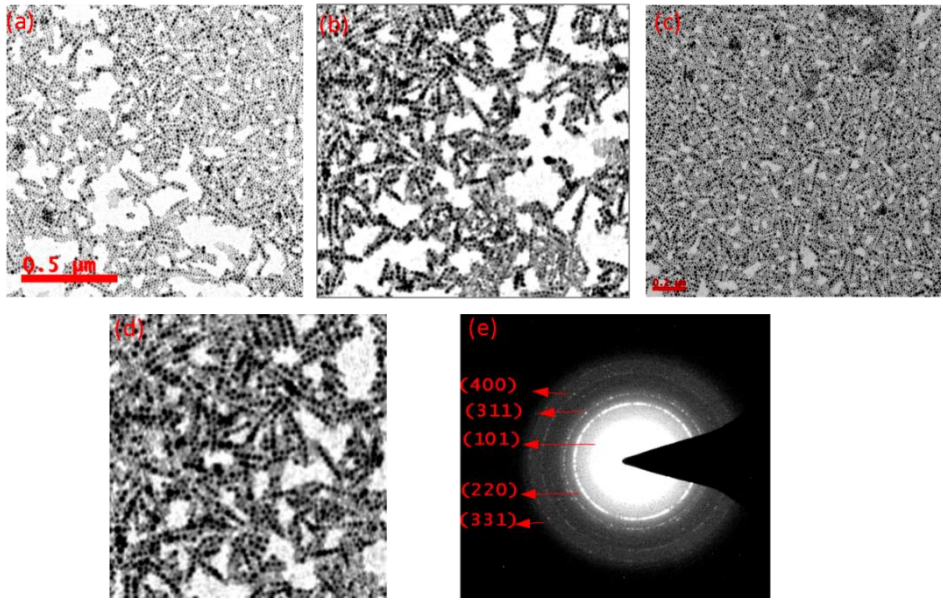

Fig. 2. Superlattice pattern type-2: shown at different magnifications. The formation of type-2 pattern has been involved with its corresponding SAED pattern.

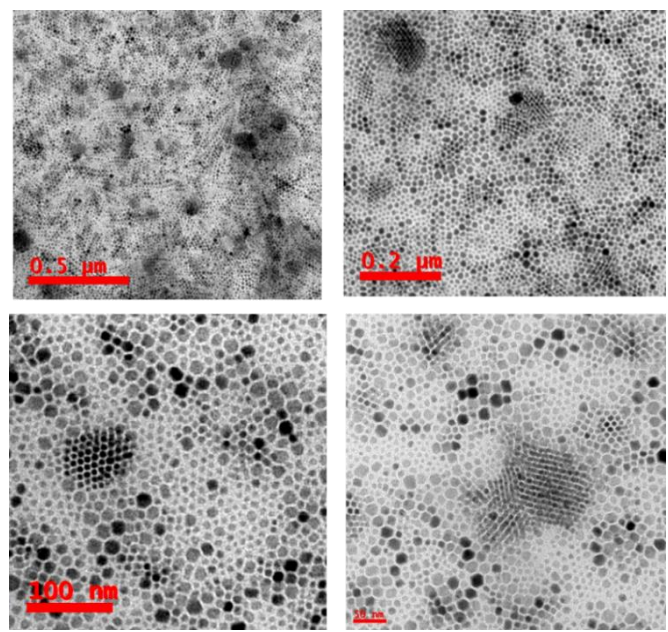

Fig.3. Superlattice pattern type 3 shown at different magnifications.

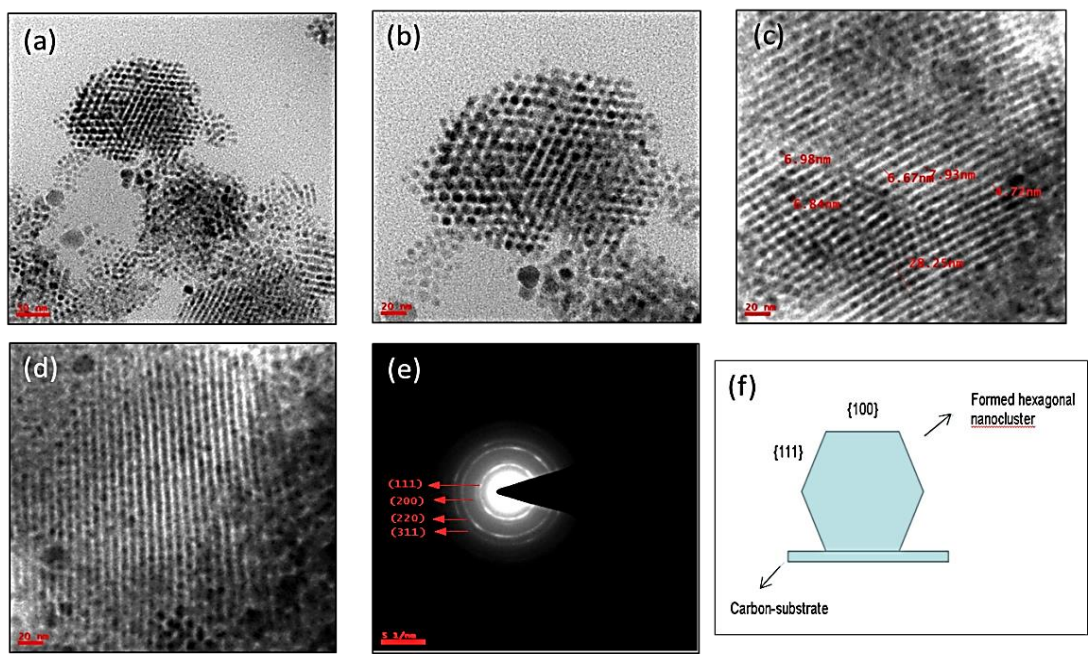

Fig. 4. TEM micrograph of self-assembled super lattice formations of UCNPs with 15-20 mins at high reaction temperature, Fig. (a and b) for short range order SL structures at low magnification, Fig. (c and d) for long range order SL structures (20nm scale bar), Fig. (e) SAED pattern for the SL taken on (d). Fig. (f) Shows the schematic representation of the orientation and growth direction of SL with UCN-NPs.

For SL of nanomaterial, it is the structure of surface and crystal on which their chemical and physical properties depend. TEM micrographs (in Fig. 1-6) revealed the structures of those size and shape controlled self-assembled nanocrystals. TEM images for different UCN-SL colloidal samples are exhibiting the morphology of self-organized and periodic nature of synthesised materials. In self-assembled SL structure, the shape and size-controlled nanoparticles act like molecules which can be considered as building blocks to construct 2D and 3D self-assembled SL-clusters.

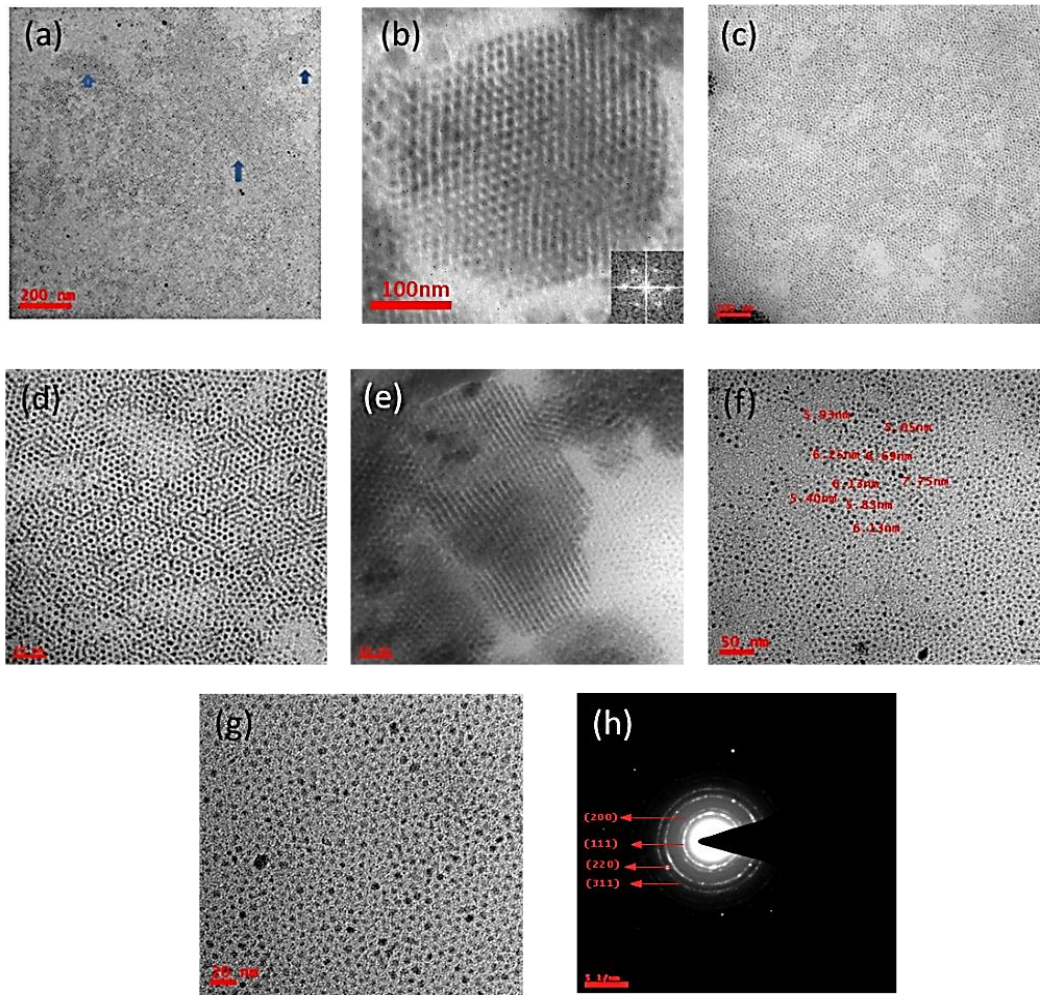

Figure 5: TEM micrographs of self-assembled formations of UCNPs formed SL with reaction time of 51 mins at high reaction temperature and at different magnification. Fig. (a) 200nm, Fig (b and c) – 100 nm, Fig. (d-f) 50 nm, Fig. (g) 20 nm, Fig. (h) SAED pattern.

With the size uniformity, morphology, orientation and order of particles in SL-structures the crystallography arrays have also been clearly identified (see Fig. 5). The organic chains of the capping agent (here oleic acid) contributed as both of protectors as well as interparticle–bonding agent with surrounding particles. The TEM micrographs revealed the formations of three different types of SL-structures synthesized from various samples.

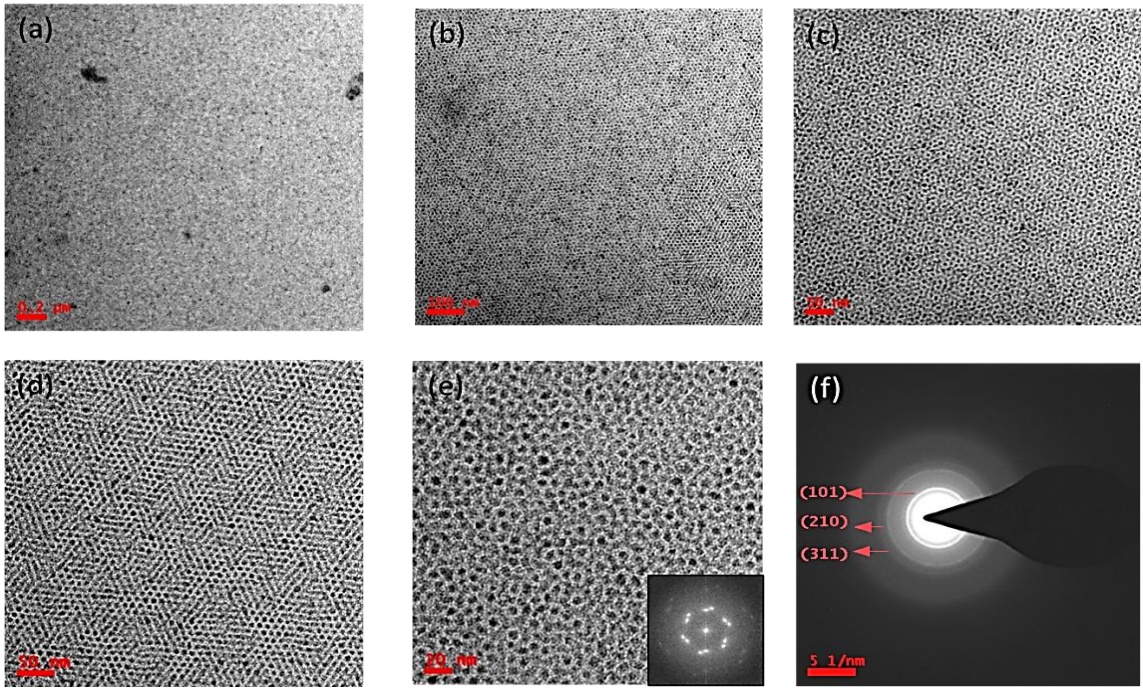

Figure 6: Shows the TEM micrographs for the superlattice structure prepared with different sizes of particles: Fig. (a) 200 nm, (b) 100 nm, (c and d) 50 nm, (e) 20 nm. Fig. (f) shows SAED pattern, Fig. (c, d and e) for clear 3D assembly with complete hexagon-type of arrangement.

The self-assembled SL (SAM-SL) formations of synthesized UCNPs prepared with different reaction conditions are shown in Fig. 4, Fig. 5 and Fig. 6. In Fig. 4, formation of SAM-SL UCNPs is shown which was obtained at very minimum reaction time (15-20 mins). Fig. 5 shows the SAM-SL which was obtained at high reaction temperature (300 ºC), for which the reaction was continued for ~51 mins. Fig. 6, SAM-SL UCNPs structure was formed by the reaction time of 1 h 50 mins and performed at above 300 ºC.

For the first set of experiment (namely exset-1, maintained 15-20 mins at high reaction temperature ~300ºC), the TEM images shown in Fig. 4 revealed the formation of self-assembled structures having [100] and [111] orientation which exhibited the formation of organized SL structure. For this sample a number of discrete nanoclusters have also been formed. The shape of each SL arrangement is almost hexagonal, that are formed due to the arrangement of nanoparticles arrays. On nanocluster-surface the distance between two adjacent parallel lines of the particles is measured to be 5 nm, which is uniform throughout the structure. Considering a single nanocluster, the particles are found to be monodispersed and they are uniform in size. In each nanocluster, the average diameter of the UCNPs is calculated to be 7-8 nm in diameter. The orientation of the particles along the surfaces of the hexagonal shaped nano-cluster was confirmed from Figure 4(b), which suggests for the formation of {100} and {111} facets on substrate and having thermodynamically unstable crystal growth as it is shown with a schematic representation in Fig. 4(f). Additionally, Fig. 4(c) and 4(d) reveal the formations of UCN-SL (as looks like disk shaped), where the UCN-particles are situated at a lower level of a disk-layer and constructing mostly (111) planes along the surface of a cluster though (100) planes.

For the second set of experiment (namely Exset-2, reaction performed for 51 mins at reaction temperature of 300$^{0}$C), TEM micrograph (Fig. 5) revealed very interesting results. During experiment, precursors were heated for a longer time compared to the previously formed 3D- assembled nano disk (check 3D/2D-self-assembled SLs) (Fig. 4). For this type of self-assembly, unstable SL-arrangements are formed as shown in Fig. 5f and 5g. Further, hexagonal arrangement nanoparticles were observed (Fig. 5d) along with a beautiful rod-like cluster arrangement with parallel assembly (Fig. 5e) is also identified. It can be noted that the monodispersed particles occupied most of the places of the arrangement. It is also noted that the self-assembled structure was broken into a separate monodispersed particles as it was heated to an elevated temperature. Further, it is noticed that the arrangement is a perfect hexagonal arrangement with uniform size of particle having diameter of ~ 50 nm (see Fig. 5d).

Next, the materials were heated for a longer period of time. However, the reaction temperature with corresponding reaction time was not enough to break all such nanoclusters completely into single monodisperse particles but many parts of them were able to overcome the energy barrier and they separated into discrete single particles. As a result, a few numbers of particles of the sample remained as self-assembled clusters and some of the particles get converted into discrete monodispersed state of the particles.

To analyse this interesting phenomenon, we increased the reaction time to 1 h 50 mins (for a third experimental set up; namely Exset-3) at 300 ºC. TEM images for the sample of Exset-3 are shown in Fig.6 which exhibited a hexagonal shaped arrangement consisting of total of 7 (seven) number of particles of which one particle is situated in the middle surrounded by another six similar sized of particles. This type of arrangement was observed up to a very long-range-order. Further, TEM images (in Fig. 6), depict clearly that most of the 2D-superlattices (as shown in Fig. 5) have been converted into a 3D-superlattice arrangement. This beautiful 3D arrangement occurred by a number of monodispersed particles of size 5 nm – 9 nm in diameter. Thus, from TEM images it is clear that these types of arrangements could be more useful to consider the shape-symmetry of as-synthesized nanocrystals and especially 3D objects with 2D projected image. The key parameters to analyse the packing between the particles are the adjustable ratio between particle-size and interparticle distance in the presence of passivated surfactant (here Oleic acid acts as a surfactant) whose chain length can be controlled. It can be noted that the TEM grids were prepared by placing the droplets of NaYF4:$Yb^{3+}$; $Er^{3+}$ superlattice colloidal solution and then it was dried properly under ambient temperature (25 ºC). Although the drying time was different for different sets of three samples however drying temperature was fixed at ~25 ºC and 10 μL of concentrated colloidal solution was taken on TEM grid and dried for 20-25 mins. Fig. 5 shows clearly a beautiful 2D monolayer self-assembled superlattice-UCNPs structure was formed in this process. It is also noticed that the particle-packing order for this type of superlattice is of short-range order as indicated by the arrowhead's region (in Fig. 5a).

Further, different parts of the sample have been examined through TEM and it confirms the unique self-assembly formation of 3D-SAM-SL-UCNPs as it is shown in Fig. 6 (a-e). From Fig. 6c and 6e, clearly identifies a perfect hexagonal shaped arrangement of the nanoparticles with one particle at the centre surrounded by another six particles and it exhibited six-fold symmetry.

Apart from the above said self-assembly superlattice phenomenon, few different types of defects have also been identified (from the TEM micrographs) in the superlattice structures. From TEM micrographs it is further noticed that for both first set (Exset-1) (Fig. 4) and second set (Exset-2) of samples (Fig.5), we have traced locations in short-range order arrangements, where the planar defects such as twinning effects along with the stacking faults like conventional crystal structure have taken place between UCN-particles of the superlattice arrangements. In the twinning defect a plane is shared by two sub-grains in such a way that one grain can be considered the mirror image of the other. Although in the present superlattice materials this twinning effect is observed with less number but still their presence owing to less volume and surface energies of smaller particles introduce such planar defects in the self-assembled superlattice arrangement.

The other type of planar defect such as Stacking fault has also been observed as it is shown in Fig. 5c, which appeared by the missing of the vertical plane (say-plane-c) between two hexagon-shaped nanoparticles.

On the other hand, the 3D self-assembly superlattice formation as shown in Fig. 6 is completely different and is very interesting compared to the type shown in Fig. 5. As shown in Fig. 6, this superlattice structure is a closely packed single layer structure consisting of UCNPs of size 6-7nm in diameter (see Fig. 6 with different magnifications such as scale bar 200nm, 100nm, 50nm and 20nm). A hexagonal-type of arrangement of the particles has been occurred (in Fig. 6c, 6d and 6e), which is a three-layer assembly. A 3D-assembled SL structure with multilayers is observed in the TEM image as it is shown Fig. 6d. Following slow evaporation of carrier solvent, a 3D-SL (Fig. 6d) was obtained with hexagonal type of arrangements sharing 12 particles with each other. Thus, the 3D-SL as shown in Fig.6 indicates that the nanoparticles are closely packed into superlattice arrangements and they are almost free from any type of defect and forms a defect-free 3D-SL structure. In conclusion, the as-synthesized SAM-SL- UCN nano crystals had tendency to self-assembled into both 2D and 3D structures owing to their uniform size of particles.

Further, the solid state crystal structure for the superlattice samples that are shown in TEM micrographs for Exset-1, Exset-2, and Exset-3 in Fig. 4, Fig. 5 and Fig. 6, respectively, confirm the formations of cubic crystalline structure with the orientation of the crystalline planes such as [(111), (200), (220) and (311)], cubic [(111), (200), (220) and (311)] and mostly hexagonal phases [(101) ,(210) and (311)] [43–47], respectively. The appearances of diffraction planes from TEM-SAED pattern for each set of samples are matching well with their resulted obtained from the XRD patterns as it is shown in the supplementary information (Fig. S1). These results have also been confirmed from the Raman analysis. The Raman spectra of three different synthesized samples are acquired and confirming the presence of different phases in synthesized sample (Figure not shown). The resulting vibrational modes that appeared are matching well with the previously reported results [48–52]. Additional bands present is due to the capping agent (oleic acid) [53].

The upconversion fluorescence spectra of three differently synthesized SAM-SL UCNPs are shown in Fig. 7 (a), 7(b) and 7(c), for Exset-1, Exset-2, and eExset-3, respectively. For the samples Exset-1 and Exset-3, fluorescence spectra are observed in both the UV and visible regions. The intensity of the emissions bands that appeared in the visible region is much higher than the emission bands that appeared in the UV region. As evidence, the full range emission bands are shown in Fig. 7(a-i) and 7(c-i) and UV emission bands are shown in Fig. 7(a-ii) and 7(c-ii) for the samples Exset-1 and Exset-3, respectively. The maximum

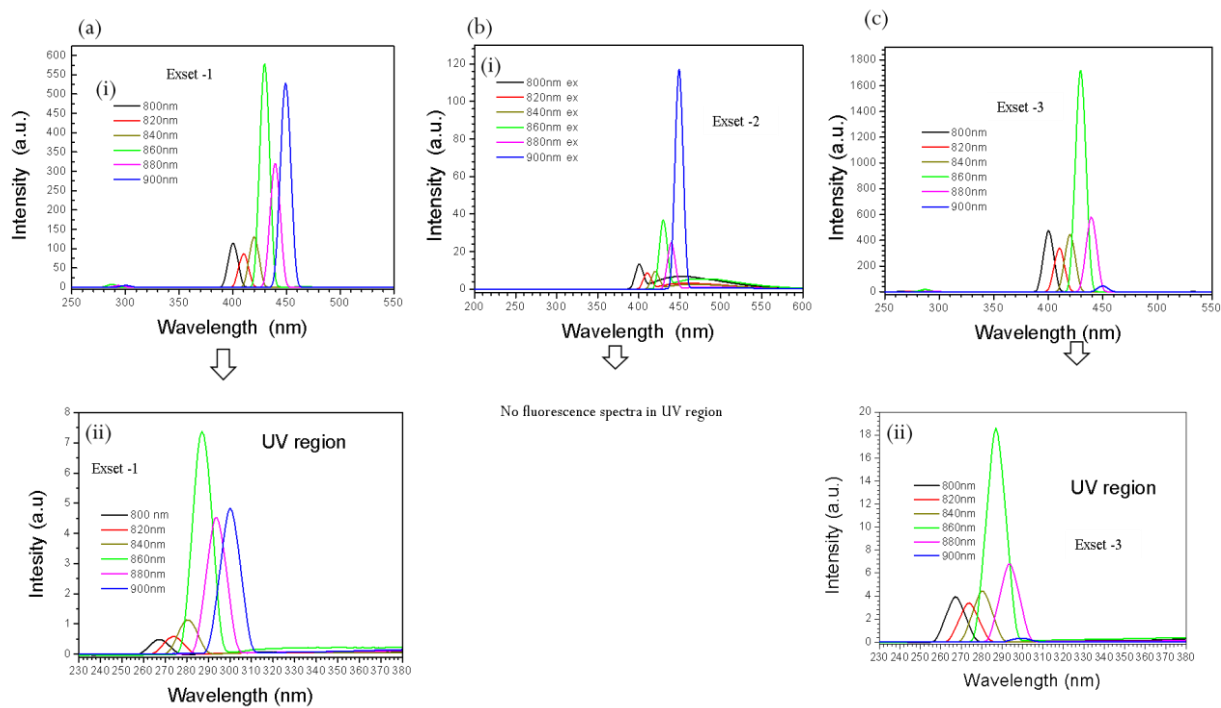

Fig. 7 (a-b) Fluorescence spectra recorded at room temperature with varying excitation wavelengths from 800nm to 900nm. Fluorescence emission spectra shown for (a) Exset-1: (i) full region and (ii) UV region; (b) Exset-2 (only full region is shown (note: no fluorescence emission occurred in UV region); (c) Exset-3: (i) full region and (ii) UV region emission bands appeared for the samples e.g., Exset-1 and Exset-3 at the spectral position (wave length) of 267 nm, 274 nm, 280 nm, 287 nm, 294 nm and 300 nm in UV region; and at 400 nm, 410 nm, 420 nm 430 nm, 439 nm and 449 nm in visible region under the 800 nm, 820 nm, 840 nm, 860 nm, 880 nm, and 900 nm excitation wavelength (in the NIR range), respectively. Interestingly, clear red siftings are observed in both the samples (i.e., for the Exset-1 and Exset-3) with shifting of bands towards higher excitation wavelength regions. It is noticed that the highest intense emission bands observed under the NIR excitation of 860 nm (for both in UV and visible regions). For the other sample (i.e., for Exset-2) emission bands are observed only in visible region with appearing highest intense emissions once it was excited with the 900 nm NIR source and the intense emission bands appeared at $\lambda$ = 401 nm, 410 nm, 421 nm, 430 nm, 440 nm and 449 nm with similar red shifting as the NIR excitation wavelength were increased. However, the fluorescence intensities of the emission band spectra for the sample Exset-3 are much higher and about 2.9 times for UV region and 2.5 times in the visible region compared to the sample Exset-1 and about 14.7 times higher in visible region compared to the Exset-2 sample (calculation was done with considering the highest intense band positions in each case). In visible region all the emission spectra fall in between λ = 400 nm – 450 nm, satisfying the blue emissions and the other emission bands appeared in the UV region. The corresponding energy band diagram is shown in Fig. 8(d) which is following a continuous direct energy transfer from $Yb^{3+}$ and $Er^{3+}$ ions. Due to the energy transition following the path: $2F_{7/2}(Yb^{3+})$ - $2F_{5/2}(Yb^{3+})$ -$2H_{9/2}$ $(Er^{3+})$ – $4I_{15/2}(Er^{3+})$, blue emissions occurred and due to the energy transitions $2F_{7/2}(Yb^{3+})$ - $2F_{5/2}(Yb^{3+})$ - $2H_{9/2}$ $(Er^{3+})$ – $4G_{11/2}$ $(Er^{3+})$ - $4I_{15/2}(Er^{3+})$, the less intense UV emissions occurred. These intense visible blue emissions occurred from the as-prepared colloidal SAM-SL-UCNPs solutions and is shown in Fig. 8(e).

**Upconversion luminescence: 140-femtosecond laser pulses (80 MHz repetition rate) under 950-990nm NIR excitation sources**

High energy upconversion luminescence was observed due to the interactions of three SAM-SL-UCNPs colloidal sample-solution (Exset-1, Exset-2 and Exset-3) with ultrafast pulsed obtained from a 140-femtosecond laser under different NIR wavelength tunable between 950-990nm range. The emission from upconversion process observed between UV and visible region in between 200 nm -270 nm and 500 nm -680 nm range, respectively. Fig. 8 clearly shows a remarkable shift of the emitted bands in the UV region, however, in visible region no such shift has been observed. Intense band positions are located at 207 nm, 215 nm, 225 nm, 233 nm, 241 nm correspond to 950 nm, 960 nm, 970 nm, 980 nm and 990 nm of NIR excitation for $2^{nd}$ sample (Exset-2). Whereas, no such distinct bands appeared under 970 nm and 980 nm NIR diode laser sources. It is further noticed that the corresponding red shifting occurred by 8 nm, 10 nm, 8 nm and 8 nm with increasing the excitation wavelengths from 950nm towards 990nm, respectively, as shown in Fig 10(a-c). To compare the energy transfer, the schematic of the corresponding energy level diagram obtained is shown in Fig. 8 (d) for intense visible blue emissions under NIR excitation sources (800-900nm) and Fig 10(f) for the associated energy level diagram for different SAM-SL-UCNPs under femtosecond laser source.

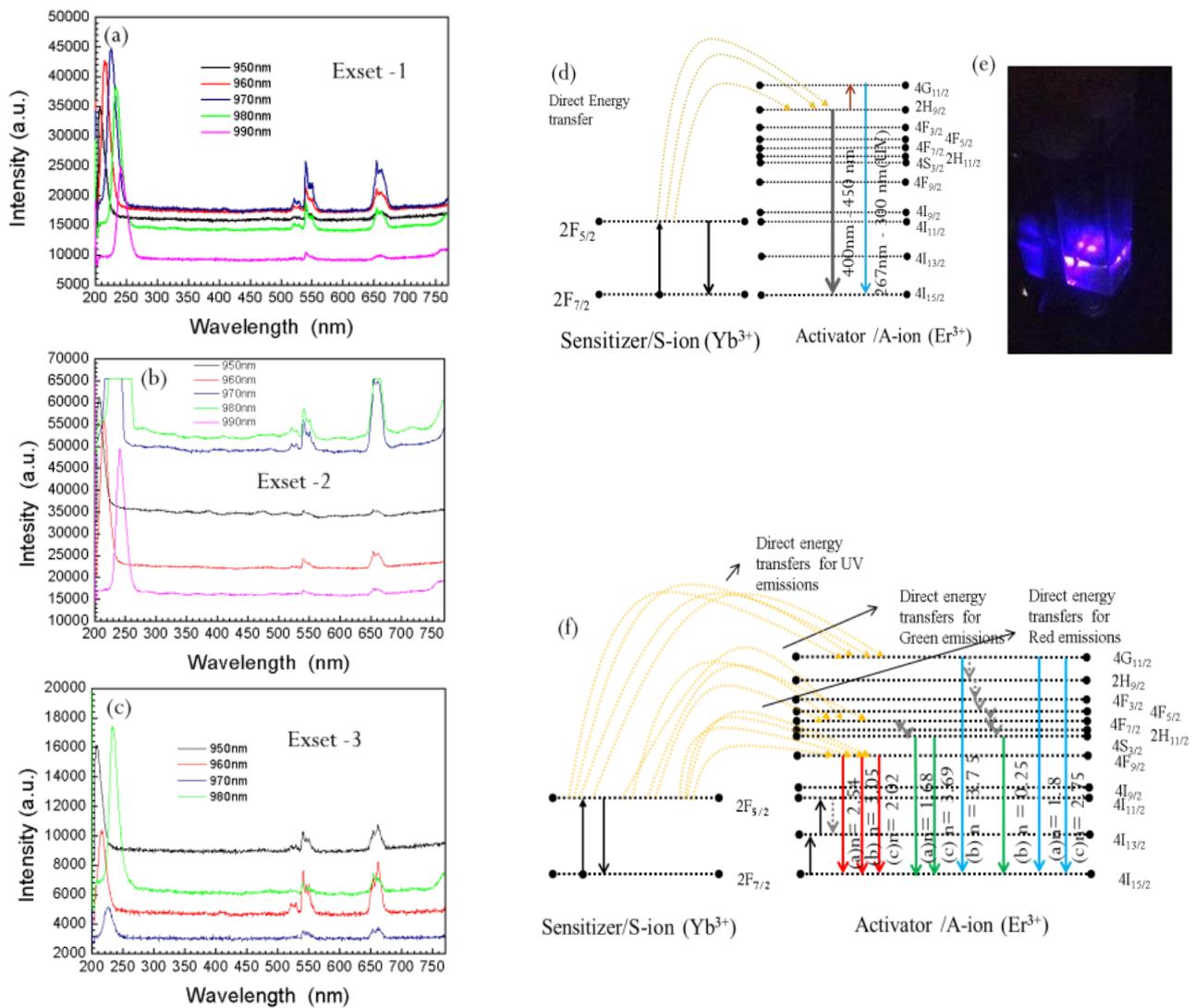

Fig. 8. Fs-laser based upconversion spectra for (a) Exset-1, (b) Exset-2, and (c) Exset 3) and (d) energy level diagram for synthesized SAM-SL–UCNPs, Fig. (e)observed intense visible blue emissions of the sample under the CW NIR excitation sources (800-900nm). Fig. (f) Associated energy level diagram for different SAM-SL-UCNPs under femtosecond laser source.

However, there is no considerable extent of shifting of bands observed for three different samples rather than change in the intensity of emission bands. Based on the baseline corrections of intensities under different pump wavelengths, the "n" values have been calculated using power law equation (1) [54–58],

$$I_{UC} \sim I_{FS}^{n} \quad \ldots \ldots \ldots \quad (1)$$

where, $I_{UC}$ = luminescence intensity,

and, $I_{Fs}$ = laser excitation power,

'n' represents the number of absorbed photons per photon emitted under the fs-laser excitation power ($I_{Fs}$) and this has been calculated from the slope of log ($I_{UC}$) versus log ($I_{Fs}$) plots (Fig.11) and their values are mentioned inside the Fig. 9(a), Fig. 9(b), and Fig. 9(c) for the UV, green and red emissions, respectively for three different samples. In UV regions, the 'n' values have been evaluated to be 1.8, 3.75 and 2.75; in green visible regions the values obtained are 1.68, 0.25 and 3.69; and for red emissions the values are found to be 2.54, 3.05 and 2.02, which are satisfying the conditions for a direct energy transfer from $Yb^{3+}$ (S-ion) to $Er^{3+}$ (A-ion) following the path: $Yb^{3+}$ ($2F_{5/2}$) - $Er^{3+}$ ($4G_{11/2}$, $4F_{7/2}$, $4F_{9/2}$).

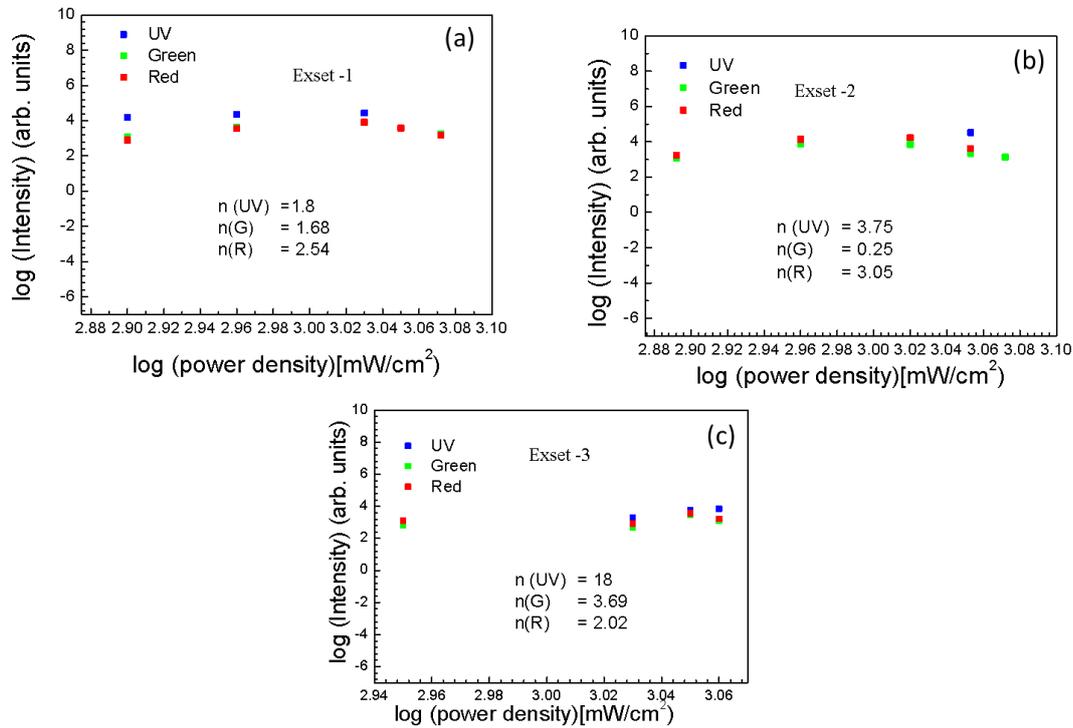

Fig. 9 Shows the plots for log ($I_{UC}$) versus log ($I_{Fs}$): the "n" –values have been calculated from fs-laser power dependent formula (eqn. 1) and have been derived in deep UV, green and red visible regions of the emission spectra for three samples (a) Exset-1, (b) Exset-2, and (c) Exset-3, respectively.

It is further found out that the number of absorbed photons per photon emitted, i.e., "n" values for all the sample are ~ falling close to 2 or n >2 which further confirmed that minimum a pair of photons are transferred from the S-ions (sensitizers) to the A-ions (activators) for UV, green and red emissions. From Fig. 9(b), it is confirmed that the 'n' value of 0.25 is the evidence for the green emission for the sample 2 (Exset-2). Further it can be elucidated that this green emission occurred not due to any direct energy transfer. However, it has been occurred due to cross relaxations between energy levels. Based on these results, an energy level diagram has been drawn to represent the energy transfer mechanisms involved in this and are responsible for the different emissions (Fig. 8(f)).

In Fig. 8(f), the corresponding "n"- values are also associated with the different emissions under three different sets of samples/experiments (which is the associated energy level diagram for different SL-UCNPs under femtosecond laser sources. Further the energy transition bands, 'n' and emission colours obtained during fs-laser interaction have been shown in Table-1 for a clear understanding of the energy transfer processes. Further, in Fig. S2 the dependence of luminescence intensity on Fs-Laser excitation power for three SL-UCNPs in colloidal solutions in (i) deep-UV, (ii) Green(G) (c) Red (R) emission, respectively. "n" value determined for the deep-UV, G and R emissions for three SL-UCNPs Expset-1, Expset-2 and Expeset-3, respectively and their "n" values also mentioned in the respective plots.

*Table 1: Shows "n" values in three different samples along with their corresponding energy transfer.*

| Parameters | Exset-1 (Designated as "a" in Fig. 9) | Exset-2 (Designated as "b" in Fig. 9) | Exset-3 (Designated as "c" in Fig. 9) |
|---|---|---|---|
| "n"- values with their corresponding emissions | 1.8 (UV emission), 1.68(G emission), 2.54 (R emission), | 3.75 (UV emission), 0.25 (G emission), 3.05(R emission), | 18 (UV emission), 3.69 (G emission), 2.02(R emission), |

| Corresponding energy transfers (UV, Green, Red) | UV emission = $2F_{7/2}$-$2F_{5/2}$ – $4G_{11/2}$ - $4I_{15/2}$ <br><br> G emission = $2F_{7/2}$-$2F_{5/2}$ – $4F_{7/2}$- $2H_{11/2}$-$4S_{3/2}$- $4I_{15/2}$ <br><br> R emission = $2F_{7/2}$-$2F_{5/2}$ – $4F_{9/2}$ – $4I_{15/2}$ | UV emission = $2F_{7/2}$-$2F_{5/2}$ – $4G_{11/2}$ - $4I_{15/2}$ <br><br> G emission (much quenched value) = $2F_{7/2}$-$2F_{5/2}$ – $4G_{11/2}$ - $2H_{9/2}$ - $4F_{3/2}$- $4F_{5/2}$ - $4F_{7/2}$- $2H_{11/2}$- $4S_{3/2}$- $4I_{15/2}$ (More no. of non-radiative relaxations present) <br><br> R emission = $2F_{7/2}$-$2F_{5/2}$ – $4F_{9/2}$ – $4I_{15/2}$ | UV emission = $2F_{7/2}$-$2F_{5/2}$ – $4G_{11/2}$ - $4I_{15/2}$ <br><br> G emission=$2F_{7/2}$ – $2F_{5/2}$ – $4F_{7/2}$- $2H_{11/2}$-$4S_{3/2}$- $4I_{15/2}$ <br><br> R emission = $2F_{7/2}$-$2F_{5/2}$ – $4F_{9/2}$ – $4I_{15/2}$ |
|---|---|---|---|

Baseline corrected intensities of different emission bands have been considered for intensity ratio plots as shown in Fig. 10. However, Figs. 10(a), 10(b) and 10(c) represent the intensity ratio plots for UV-emissions/Green emissions (UV/G), UV-emissions/Red-emissions (UV/R) and Red-emissions/Green-emissions (R/G) for three SAM-SL-UCNPs colloidal samples, respectively, which have been calculated under the different incident pump wavelengths. The efficient populations of photons in different energy levels could be the possible reason of different intense emission bands under NIR sources (950 nm-990 nm excitation wavelengths). Further, the Table S1 shows the intensity ratio values in UV and visible region under different excitation wavelengths. Further, the Quantum yield values (Fig. 11) have been calculated for different superlattice samples and they are shown in Table-2 in Table 3.

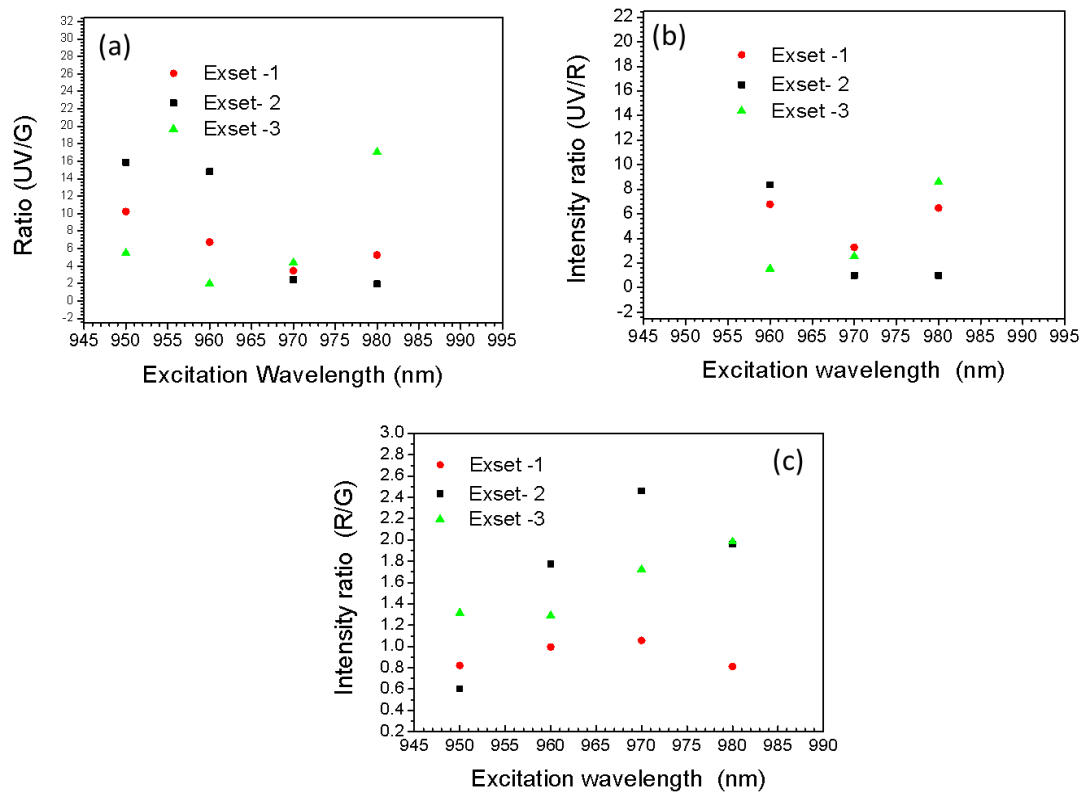

Fig. 10. Plots for the intensity ratios of different emissions of three SAM-SL-UCNPs colloidal solutions with different excitation radiations, (a) intensity ratio for (UV/Green) emission, (b) intensity ratio for (UV/Red) emission, and (c) intensity ratio for (red/green) emission, respectively.

*Table 2:* Shows the calculated Quantum Yield (QY) for the different samples.

| Samples | Quantum Yield (QY) | |
|---|---|---|
| | UV region | Vis- region |
| Exset-1 | 0.032 %-0.22% | 0.076%-0.22% |
| Exset-2 | 0.054% - 1.44% | 0.124% - 1.44% |
| Exset-3 | 0.08% - 1.17% | 0.19% - 1.17% |

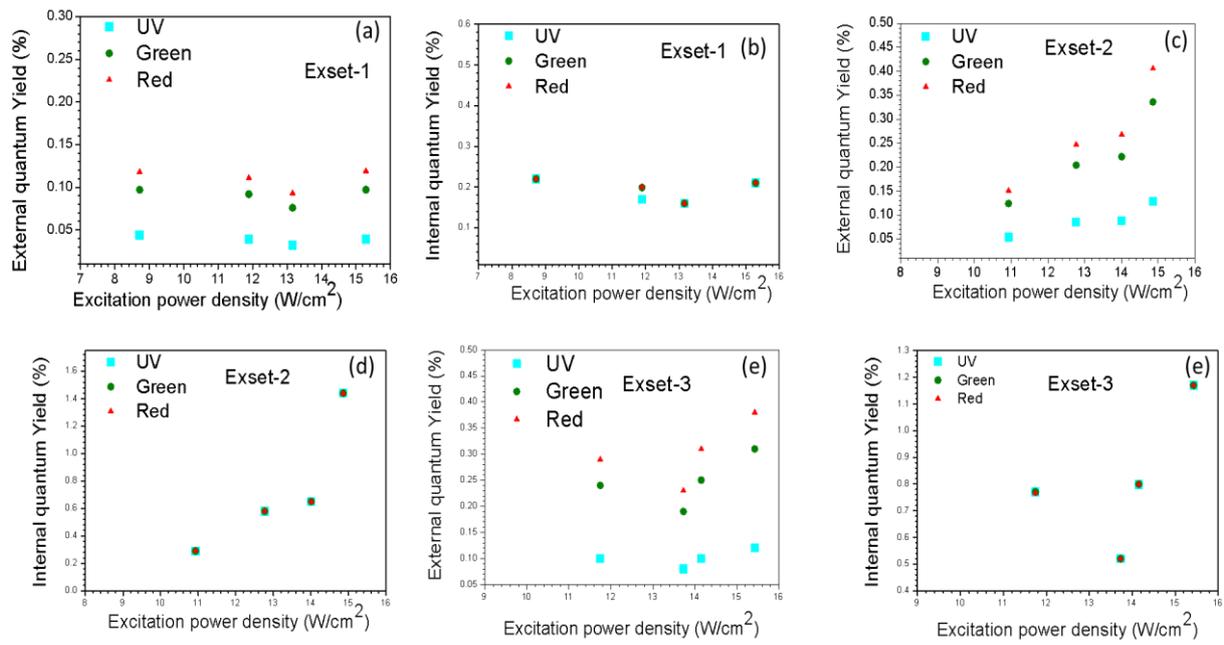

Fig. 11. Shows the results for External and Internal Quantum yields of Exset-1 ((a) and (b)), Exset-2 ((c) and (d)) and for Exset-3 ((e) and (f)), respectively.

*Table 3*: Show the Quantum yield (Q.Y.) values calculated for three different sets of samples. Here both the External Q.Y. values and Internal Q.Y. values have been calculated.

| Q.Y. values | Exset-1 | Exset-2 | Exset-3 |
|---|---|---|---|
| External Q.Y. values | **UV emission=**<br>0.039% (960 nm excitation)<br>0.032% (970 nm excitation)<br>0.039% (980 nm excitation)<br>0.044% (990 nm excitation) | **UV emission=**<br>0.129% (950 nm excitation)<br>0.088% (960 nm excitation)<br>0.085% (970 nm excitation)<br>0.054% (980 nm excitation) | **UV emission=**<br>0.12% (950 nm excitation)<br>0.099% (960 nm excitation)<br>0.08% (970 nm excitation)<br>0.103% (980 nm excitation) |
| | **G emission=**<br>0.097% (960nm excitation)<br>0.076% (970nm excitation)<br>0.092% (980nm excitation)<br>0.097% (990nm excitation) | **G emission=**<br>0.336% (950nm excitation)<br>0.222% (960nm excitation)<br>0.204% (970nm excitation)<br>0.124% (980nm excitation) | **G emission=**<br>0.31% (950nm excitation)<br>0.25% (960nm excitation)<br>0.19% (970nm excitation)<br>0.24% (980nm excitation) |
| | **R emission=**<br>0.119% (960nm excitation)<br>0.093% (970nm excitation)<br>0.111% (980nm excitation)<br>0.118% (990nm excitation) | **R emission=**<br>0.406% (950nm excitation)<br>0.268% (960nm excitation)<br>0.247% (970nm excitation)<br>0.151% (980nm excitation) | **R emission=**<br>0.38% (950nm excitation)<br>0.31% (960nm excitation)<br>0.23% (970nm excitation)<br>0.29% (980nm excitation) |
| Internal Q.Y. values | **UV emission=**<br>0.21% (960nm excitation)<br>0.16% (970nm excitation)<br>0.17% (980nm excitation)<br>0.22% (990nm excitation) | **UV emission=**<br>1.44% (950nm excitation)<br>0.65% (960nm excitation)<br>0.58% (970nm excitation)<br>0.29% (980nm excitation) | **UV emission=**<br>1.17% (950nm excitation)<br>0.798% (960nm excitation)<br>0.52% (970nm excitation)<br>0.77% (980nm excitation) |
| | **G emission=**<br>0.21% (960nm excitation)<br>0.16% (970nm excitation)<br>0.19% (980nm excitation)<br>0.22% (990nm excitation) | **G emission=**<br>1.44% (950nm excitation)<br>0.65% (960nm excitation)<br>0.58% (970nm excitation)<br>0.29% (980nm excitation) | **G emission=**<br>1.17% (950nm excitation)<br>0.799% (960nm excitation)<br>0.52% (970nm excitation)<br>0.77% (980nm excitation) |
| | **R emission=**<br>0.21% (960nm excitation)<br>0.16% (970nm excitation)<br>0.20% (980nm excitation)<br>0.22% (990nm excitation) | **R emission=**<br>1.44% (950nm excitation)<br>0.65% (960nm excitation)<br>0.58% (970nm excitation)<br>0.29% (980nm excitation) | **R emission=**<br>1.17% (950nm excitation)<br>0.799% (960nm excitation)<br>0.52% (970nm excitation)<br>0.766% (980nm excitation) |

**Summary and conclusions:** Supper lattice UCNPs with different self-assembly arrangements have been synthesized in this work as mentioned in the experimental section. The synthesized SAM-SL-UCNPs appeared as an autonomous organization of UCNPs without any kind of intervention or using any interfacial assembly method separately as reported earlier [59]. The self-assembled superlattice formations can be considered as static-self-assembly (which are stable after formations). A one-pot chemical synthesis approach was applied for the preparation of these UCNPs with narrow particle size distribution (particle size 5-9 nm). HRTEM images have provided direct proof of such formations. The synthesized SAM-SL-UCNPs are stable and did not proclaim any oxidation. The nearly spherical UCNPs had a great tendency to form 2D and 3D superlattice structures. The formations of such 2D and 3D SL structure patterns controlling ambient reaction parameters must be having strong importance in several practical applications such as plasmonic metamaterials [60–64], solar cells [65, 66], due to their nanoscale-architectures and can also open new windows in application bioimaging [67, 68] and photodynamic therapy [69, 70] and many other fields. The morphologically controlled SAM-SL-UCNPs with bright fluorescence introducing high intense blue emissions and generation of high energy upconversion under femtosecond laser treatment make them unique. The applications of upconversion emissions under different femtosecond laser irradiations are well known [71–75]. Now, the combination of self-assembled morphology along with the strong upconversion emissions (strong fluorescence under 800 nm-900 nm NIR irradiations and strong luminescence mainly in UV region under 950 nm-990 nm NIR irradiations) could open strong applications in biomedical fields which is still under the investigation.

**ASSOCIATED CONTENT:** Supporting Information: Supporting information (SI) are available

**AUTHOR INFORMATION:** Corresponding author: E-mail: paik.bme@iitbhu.ac.in, pradip.paik@gmail.com, akcsp@uohyd.ernet.in, akcphys@gmail.com

**AUTHOR CONTRIBUTIONS:** The manuscript was written through the contributions of all the authors and all authors l have given approval to the final version of the manuscript to submit for publication.

**Conflict of Interest.** There is no conflict of interest

ACKNOWLEDGMENT

Authors acknowledge the financial support awarded to P. Paik by DST-Nanomission, India (Ref: SR/NM/NS-1005/2015), Science and Engineering Research Board (SERB), India (Ref: EEQ/2016/000040). The author A.K. Chaudhary acknowledge the funding agency Defense Research and Development Organization (DRDO), Ministry of Defense, Govt. of India for the financial support under the grant No. DRDO/18/1801/2016/01038: ACRHEM-Phase-III

[37] C. Zhang, L. Yang, J. Zhao, B. Liu, M.-Y. Han, and Z. Zhang, "White-Light Emission from an Integrated Upconversion Nanostructure: Toward Multicolor Displays Modulated by Laser Power," *Angew. Chemie Int. Ed.*, vol. 54, no. 39, pp. 11531–11535, Sep. 2015.

[38] F. Wang and X. Liu, "Multicolor Tuning of Lanthanide-Doped Nanoparticles by Single Wavelength Excitation," *Acc. Chem. Res.*, vol. 47, no. 4, pp. 1378–1385, Apr. 2014.

[39] X. Huang *et al.*, "Realizing highly efficient multicolor tunable emissions from Tb3+ and Eu3+ co-doped CaGd2(WO4)4 phosphors via energy transfer by single ultraviolet excitation for lighting and display applications," *Dye. Pigment.*, vol. 151, pp. 202–210, 2018.

[40] J. Zhang, C. Mi, H. Wu, H. Huang, C. Mao, and S. Xu, "Synthesis of NaYF4:Yb/Er/Gd up-conversion luminescent nanoparticles and luminescence resonance energy transfer-based protein detection," *Anal. Biochem.*, vol. 421, no. 2, pp. 673–679, 2012.

[41] W. J. Kim, M. Nyk, and P. N. Prasad, "Color-coded multilayer photopatterned microstructures using lanthanide (III) ion co-doped NaYF4 nanoparticles with upconversion luminescence for possible applications in security," *Nanotechnology*, vol. 20, no. 18, p. 185301, 2009.

[42] L. Wang *et al.*, "Fluorescence Resonant Energy Transfer Biosensor Based on Upconversion-Luminescent Nanoparticles," *Angew. Chemie Int. Ed.*, vol. 44, no. 37, pp. 6054–6057, Sep. 2005.

[43] J.-C. Boyer, F. Vetrone, L. A. Cuccia, and J. A. Capobianco, "Synthesis of Colloidal Upconverting NaYF4 Nanocrystals Doped with $Er^{3+}$,$Yb^{3+}$ and $Tm^{3+}$,$Yb^{3+}$ via Thermal Decomposition of Lanthanide Trifluoroacetate Precursors," *J. Am. Chem. Soc*, vol. 128, no. 23, pp. 7444–7445, 2006.
33

**Supporting information**

# Morphologically controlled synthesis of self-assembled Upconversion superlattices and tuneable femtosecond laser interaction-based luminescence

*Monami Das Modak[c], Anil Kumar Chaudhary[b*] and Pradip Paik[a*]*

[c]School of Engineering Sciences and Technology, University of Hyderabad (UoH), Hyderabad, TS, 500 046

[b]Advanced Centre of Research in High Energy Materials, University of Hyderabad (UoH), Hyderabad, TS, 500 046

[a]School of Biomedical Engineering, Indian Institute of Technology (IIT)-BHU, Varanasi, UP, 221 005

*Corresponding author's email: paik.bme@iitbhu.ac.in, pradip.paik@gmail.com,

# Supporting Figures

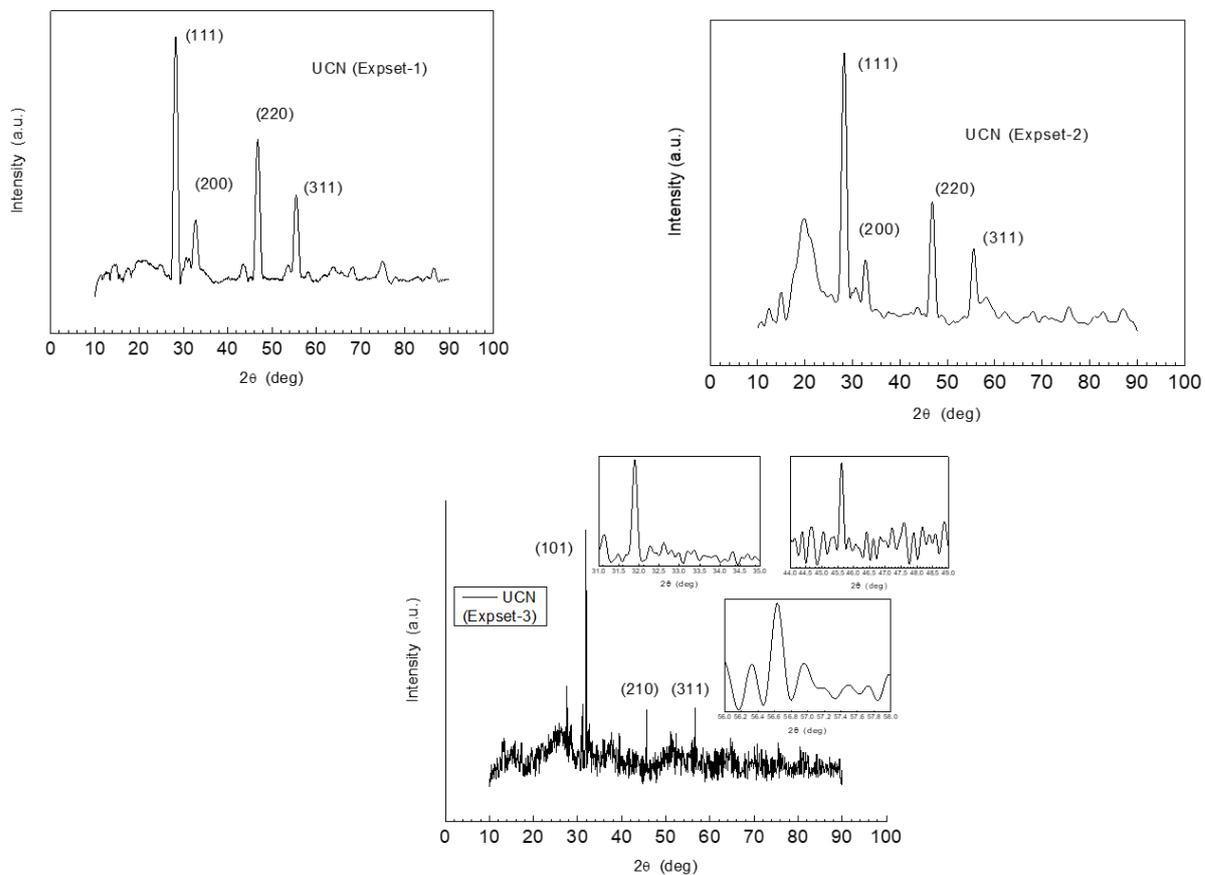

Figure S1: XRD patterned for the different set of samples self-assembled into superlattice structure.

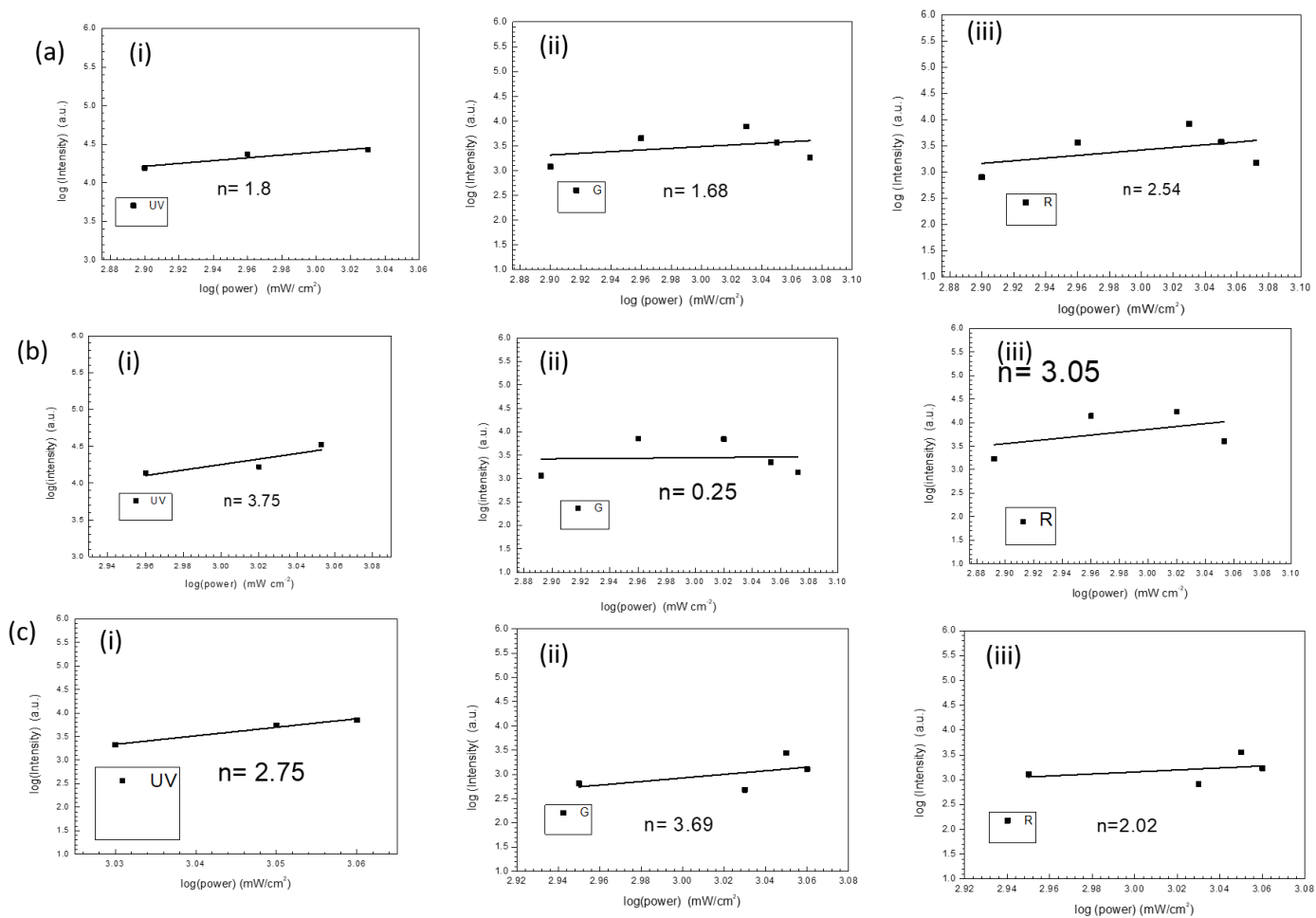

Fig. S2   Dependence of luminescence intensity on Fs-Laser excitation power for three SL-UCNPs colloidal solutions (a,b,c) in (i) deep-UV , (ii) Green(G) (c ) Red (R) emission respectively. "n" value determined for the deep-UV, G and R emissions for three SL-UCNPs Expeset-1, Expset-2 and Expe-set-3, respectively and their "n" values also mentioned in the respective plots.

**Table S1**: Tabulated form of intensity ratio values in UV and visible regions under the different excitation wavelengths.

| UV region | | | | Visible region | | |
|---|---|---|---|---|---|---|
| Ex. W.L. | | UC intensity | Ratio values | Ex. W.L. | Highest intensity | Ratio values |
| 960 | (exset-1) | 25278 | 2.27 | 970(G) | 7722 | 2.79 |
| 970 | | 26734 | 2.40 | 970(R) | 8154 | 2.28 |
| 960 | (exset-2) | 33235 | 2.99 | 980(G) | 7112 | 2.56 |
| 990 | | 33422 | 3.01 | 970(R) | 16923 | 4.74 |
| 980 | (exset-3) | 11098 | ------- | 960(G) | 2772 | ------- |
| | | | | 960(R) | 3573 | |